\def\be{\begin{equation}}
\def\ee{\end{equation}}
\def\bea{\begin{eqnarray}}
\def\eea{\end{eqnarray}}
\def\ba{\begin{array}}
\def\ea{\end{array}}
\def\nm{\nonumber \\ }

\def\sp{\qquad ; \quad }
\def\s{\ , \ }

\def\aij{\alpha_{ij}}

\def\p{\partial}
\def\pq{\partial_{\kappa}}

\def\v{\vert 0 \rangle }
\def\Ai{{\cal A}}

\def\is#1{\vert #1 \rangle}

\def\b{\beta}
 
\def\b2{\beta ^2}

\def\vf{\varphi}
\def\k{\kappa}

\def\wa2{$W\! \! A_2 \   $}
\def\wbc2{$W\! BC_2 \ $}
\def\wan{$W \! \! A_n \ $}
\def\wbcn{$W \! BC_n \ $}
\def\wg{$W\! G \ $}

\def\NPB#1#2#3{{\rm Nucl. Phys.} {\bf B#1} (#2) #3}
\def\PLB#1#2#3{{\rm Phys. Lett.} {\bf #1B} (#2) #3}

\def\IJMPA#1#2#3{{\rm Int. J. Mod. Phys.} {\bf A#1} (#2) #3}

\def\CMP#1#2#3{{\rm Commun. Math. Phys.} {\bf #1} (#2) #3}
\def\PR#1#2#3{{\rm Physics Reports } {\bf #1} (#2) #3}
\def\SCRAP#1#2#3{{\rm Sov. Sci. Rev. A. Phys.} {Vol. #1} (#2) #3}

\def\AP#1#2#3{{\rm Ann. of Phys.} {\bf #1} (#2) #3}

\documentstyle[12pt]{article}
\advance \topmargin by -\headheight
\advance \topmargin by -\headsep

\evensidemargin \oddsidemargin
\begin{document}

\title{  \begin{flushright}
\normalsize{ ITP Budapest Report No. 523 \\
hep-th//9702183 }
  \end{flushright}
\vspace{5cm}
     On the free field realization of $W\! BC_n$ algebras }
\author{ \large{ Z. Bajnok } \\ \\
\normalsize {\it Institute for Theoretical Physics }\\
\normalsize {\it  Roland E\"otv\"os   University, }\\
\normalsize {\it H-1088 Budapest, Puskin u. 5-7, Hungary}}

\def\today{February 26, 1997 }

 \maketitle

\vspace{1.5cm}

 \begin{abstract}

Defining the \wbcn algebras as the commutant of certain screening charges 
a special form for the classical generators is obtained which does not 
change under quantisation. This enables us to give explicitly the first 
few generators in a compact form for arbitrary \wbcn algebras.

\end{abstract}

\newpage

\section{ Introduction}

The first definition of the $W$-algebras was given by Fateev and 
Lukyanov in \cite{FaLuW}. They defined the classical $W$-algebras 
using the Miura transformation, consequently  the algebras were  
presented in their free field form. In order to find the quantum 
analogue they quantised the free fields and 
normal ordered the classical expressions. Then the screening 
charges, operators that commute with the $W$-generators, were identified 
since both the screening operators and the $W$-generators are necessary in
the investigation of the representation theory. Unfortunately this 
quantisation procedure works only for the $A_n$ and $D_n$ series, 
but fails to work for $C_n$ and $B_n$. To avoid this problem they modified
the screening operators and the $W$-generators simultaneously for the 
$B_n$ algebras but in doing so the correspondence to the classical 
theory is lost. 

Using an alternative definition of the quantum $W$-algebras Feigin and 
Frenkel realized \cite{FrenQG}, that most naturally one can 
define the $W$-generators and so the algebras as the commutant of  
the screening charges. They also computed the classical counterpart
and showed the relation between them: they first proved the 
existence of the classical generators then proved their survival at 
the quantum level. 

Although the existence of the quantum generators is established their 
explicit form is still unknown in the general case.  
In fact, Lukyanov and Fateev's  quantum expressions for $A_n$ and
$D_n$ still work, since in this case the polynomial form of the 
$W$-generators does not
change only the coefficients get quantum corrections. The aim of this paper     
is to find similar expressions for the missing $B_n$ and $C_n$ cases, 
ie. to find such a form for the generators at the classical level, 
which in this sense survive quantisation. Since there are applications 
where the $W$-generators are explicitly needed \cite{FuGaPeW,C_2}
we also give explicit expressions, in a very compact form, 
for the first few generators of arbitrary \wbcn algebras. (See also 
\cite{Ito} for related results). 

The paper is organised as follows:  in  Section 2 we review the 
earlier results concerning the definition and quantisation of 
$W$-algebras. Then in Section 3 we consider the classical theory and find 
a form for the generators which turns out to be very useful at 
the quantum level. In Section 4 we prove that the classical form does not 
change and give explicit results. Afterwards we conclude in Section 5.

\section{ \wg algebras as commutant of screening operators}

Usually the \wg algebras are defined as the reduction of the Kac-Moody (KM)
algebras corresponding to the principal $sl_2$ embeddings \cite{BaFeFo1}.
On can show that this is equivalent to look only for the Cartan currents 
of the KM algebra in question and find the commutant of certain 
screening charges \cite{Frenred}. 

In more detail one defines the 
\wg  algebra as a vertex operator sub-algebra of the KM algebra's Cartan 
sub-algebra. The commutation relations of the modes of the ghost-modified 
 Cartan currents \cite{TjdB,BouShou} are en-coded in
\be
\hat J^i_1\is{\hat J^j_{-1}}=\alpha_{ij}(k+h)\v =
\alpha_{ij}\k \v ,
\ee
where $k$ denotes the level and $h$ the dual Coxeter number. The $\hat J ^i$
correspond to the simple roots,$ \alpha_i$, and $\alpha _{ij}=\alpha_i\cdot
\alpha_j$ is the inner product matrix of $G$. 
The space of states is spanned by the 
negative modes of the currents and the \wg algebra is nothing but 
the only non-vanishing cohomology of the BRST charge $\sum_i Q_i$,
consisting of the sum of the screening charges, $Q_i$, 
\be
Q_i=\sum_{n\leq
0}S_{n+1}^i(\hat J^i_-) S_n^i(\hat J^i_+)
 \quad ; \quad 
\sum _{n\leq 0} S^i_n(x^i_{\pm})z^n=\prod _{m<0} e^{\pm{x^i_{{\mp}m}
 \over m }z^m} .
\ee
The kernel of $Q_i$ coincides with the kernel of 
$\oint {dz\over 2\pi i} :e^{\int ^z \hat J^i(w) dw}:$.
 
The classical limit can be associated to
 $\k \to 0$ \cite{FrenQG}. In this case the 
classical screening charges become:
\be
\tilde Q_i=-\k^{-1}\lim _{\k \to 0}Q_i=\sum_{n<
0;j}S_{n+1}^i(\hat J^i_-) \aij {\p \over \p \hat J^j_{n}}=\sum_{n<
0}S_{n+1}^i(\sum_j\aij h^j_-)\p^i_{n}\sp \p^i_{-n}={\p \over \p h^i_{-n}} ,
\ee 
where we have introduced the $\sum_i\aij h^j_n=\hat J^i_n$ variables, which 
correspond to the fundamental co-weights. Here, analogously to the quantum
case, $h^i_{-m-1}$ represents the $(m!)^{-1}\p ^mh^i(z)$ classical field. 
The classical analogue of $L_{-1}$ is  $ \p= \sum_{m< 0}                       
\sum^{dim g} _{i=1} m h^i_{-m-1}{\p^i_{m}}$, which is the operator
of the differentiation on the differential polynomials of the classical 
fields, $h^i(z)$. 

Using the fact that $\tilde Q_i$-s  satisfy the Serre relation
of the algebra Feigin and Frenkel proved that the kernel of these 
operators is spanned by elements $W_{e_i+1}$ of degree $e_i+1$ 
together with  products 
of their derivatives. Here the $e_i$-s denote the exponents of the Lie 
algebra. Modifying the proof slightly 
 one can show that taking a subset $I\subset 
\{1, \dots, n\}$, the common kernel of the 
$\tilde Q_i$ operators  for $i\in I$  consists of the $W$-generators 
of the associated sub-algebra, (whose simple roots are $ \alpha_i \s i\in I$), 
and the remaining $h^j_n\ , \ j\not \in I$  variables. 
As a consequence the kernel of $Q^{i}$ is 
generated by $L^{(i)}$ and $h^j_{n},\ j\neq i$, where 
$L^{(i)}$ is defined as 
\be
\p^2+(h_{i+1}-h_{i-1})\p
+L^{(i)}=(\p+h_{i+1}-h_i)(\p+h_i-h_{i-1}) ,
\ee
 (if $\alpha_{ij}=2\delta_{ij}-
\delta_{i-1j}-\delta_{i+1 j}$).  Here and from now on we abbreviate
$h_{-1}^i$ by $h_i$. The translation covariance of this expression 
motivates one to find the $A_n$ generators in the following form:
\be
(\p-h_n) \dots (\p +h_{i+1}-h_i)(\p+h_i-h_{i-1})
\dots(\p +h_1)=\sum_{i=0}^n W_i^{A_n} \p ^{n-i} . 
\label{ancl}
\ee
Unfortunately this translation covariance is absent in the 
the $C_n$ and $B_n$ case, for this reason we choose the following 
strategy. As a starting point we consider 
the first $n-1$ screening operators. Since they are exactly the same 
as in the $A_n$ case one knows their kernel and the 
 corresponding generators. In order to find the common kernel we 
need $L^{(n)}=h_nh_n-2h_nh_{n-1}-\p h_n$
 so we combine (\ref{ancl}) with its adjoint and obtain
the correct answer:
\be
(\p -h_1) \dots (\p+h_{n-1}-2h_n)(\p +2h_n-h_{n-1}) \dots
(\p +h_1)=\sum_i W_{2i}^{C_n}\p^{2(n-i)}+ {\rm odd \ terms} .
\label{wcn}
\ee  
One has a similar solution for $B_n$:
\be 
(\p -h_1) \dots (\p+h_{n-1}-h_n)\p (\p +h_n-h_{n-1}) \dots
(\p +h_1)=\sum_i W_{2i}^{B_n}\p^{2(n-i)+1}+{\rm even \ terms} .
\label{wbn}
\ee

The quantum case is much more involved. 
Since the 
\be
Q_i=\sum_{n\leq 0}S_{n+1}^i(\hat J^i_-) S_n^i(\hat J^i_+)=
\sum_{n\leq 0}S_{n+1}^i(\sum_j\aij h^i_-)S_n(h^i_m\to -m\k\p^i_{-m})
=-\k \tilde Q_i+\k^2 (\dots )
\ee
quantum operators
satisfy the q-Serre relations of the underlying q-deformed Lie algebra
the classical $W$-generators survive 
\cite{FrenQG}. However this means that the commutant of the subset 
of the $Q_i$ operators, in harmony with the classical theory, consists
of the $W$-generators of the associated sub-algebra and the remaining 
$h^j \s j\neq i$ variables. To find the explicit quantum expressions
one tries the classical form modified according to $L^{(i)}_{q}=L^{(i)}_{cl}(
\p\to (1-\k)\p)$.  It works for $A_n$:
\be
(\pq-h_n) \dots (\pq +h_{i+1}-h_i)(\pq+h_i-h_{i-1})
\dots(\pq +h_1)=\sum_{i=0}^n W^{A_n}_i \pq ^{n-i} \sp \pq=(1-\k)\p .
\label {anq}
\ee
In the other cases this naive quantisation does not work since one 
has to deal with the corrections coming form 
the normal ordering of more then two $h^i$-s. 

Taking into account the special form of the $Q_i$ operators (the $\k^n$ 
terms contain $n$ differentiations) one can show that each quantum 
generator has the following form \cite{Nieab}:
\be
P^{\k}=P_0+\k P_1+\dots +\k^iP_i+\dots  +\k^{k-1}P_{k-1} ,
\ee
where $P_0$ is the classical expression, eq. 
(\ref{ancl},\ref{wcn},\ref{wbn}),
 and $P_i$ contains the product of 
at most $k-i$ terms. Moreover the sum of the longest terms, terms
without derivatives, gives a Weyl invariant polynomial.

\section{Classical considerations }

We have seen in the previous section that the zeroth order term in the 
quantum expression is the classical generator. This motivates us to 
start at the classical level where we set up the notations strongly 
suggested by the quantum problems. 

Since we know how the $A_n$ generators
get quantum corrections, (\ref{anq}), we rewrite the classical generator
(\ref{wcn}) into to following form:
\be
   \sum_i W_{2i}^{C_n}\p^{2(n-i)}+{\rm odd \ terms}
  =\left ( \p^n-\p^{n-1}W_1+\dots +(-1)^n W_n\right )
\left (W_n+\dots  +W_1\p^{n-1}+\p^n \right ) .   
\ee
In order for $(\p +2h_n-h_{n-1})$ to be the same as in the $A_n$ case:
$(\p+h_n-h_{n-1}) $, we rescaled $h_n$ by a factor two. Here the $W_k$-s 
span the kernel of the $Q_1, \dots ,Q_{n-1}$ operators. Clearly $W_1=h_n$. 
Moreover if one takes the $h_1 \to 0$ limit then $W_n \to 0$ and for 
all the others we 
recover the generators related to the $A_{n-1}$ algebra. 

Unfortunately in the definition above the $W^{C_n}_{2i}$ generators 
explicitly depend on $n$. We would like to redefine them $n$-independently. 
We also need a form for the $W_{2k}^{C_n}$ generator such that 
$W_{2k}^{C_n}\to 0$ when $W_l\to 0 \s l=n,n-1, \dots k$. 
This can be achieved by the following way: define the generators by
\bea
& \left ( \p^n-\p^{n-1}W_1+\dots  +(-1)^n W_n\right )
\left (W_n+\dots   +W_1\p^{n-1}+\p^n \right )= \nm
& \p^n \p^n+\p^{n-1}W^{C_n}_2\p^{n-1}+\p^{n-2}W^{C_n}_4\p^{n-2}+
\dots +\p^{n-k}W^{C_n}_{2k}\p^{n-k}+\dots + W^{C_n}_{2n} .
\label{nwcn}
\eea
It is not completely clear at first
 that this definition is correct. Writing 
this expression into the original form (\ref{wcn}), one can see that 
at each even level (\ref{nwcn}) really 
defines the generator, however at odd levels
it is a nontrivial statement. In order to prove it first we 
note that 
\be
W^{C_n}_{2n}=\left ( \p^n-\p^{n-1}W_1+\dots +(-1)^n W_n\right )W_n .
\ee
This shows that if we take the $h_1\to 0$ limit then $W_{2n}^{C_n}\to 0$
while for all the other generators we have  $W_{2k}^{C_n} \to 
W_{2k}^{C_{n-1}}$. This 
indicates that the  $W_{2k}^{C_n}$ generators share the 
property needed above, ie. $W_{2k}^{C_n}\to 0 $ when $ W_l \to 0 \s
l=n, \dots ,k$. It is also clear that the generators are $n$-independent. 
Now the terms at odd level  
have to be in the kernel of all the screening charges. This means 
that they have to be a linear combination of the derivatives of
the $W_{2k}^{C_n}$ generators defined at
even levels. However one can determine the coefficients of this 
combination from the $W_kW_k$ term of $W_{2k}^{C_n}$, which are exactly
what is needed. 

Summarizing the classical generator has the following form:
\be 
W^{C_n}_{2k}=\sum _{i,j,l,m} c^{lm}_{ij}\p^iW_l\p^jW_m  ,
\label{clgen}
\ee
where $i+j+l+m=2k$ and $l\leq k \leq m$.  
  
Here we list the first few generators since we will need their explicit
form at the quantum level.  
\bea
W^{C_n}_2&=&2W_2-\left (W_1\right )^2+\p W_1 \nm
W^{C_n}_4&=&2W_4+3\p W_3-2W_1W_3+W_2W_2-(\p W_1)W_2-W_1\p W_2
+ \p ^2 W_2 \\
W^{C_n}_6&=&2W_6-2W_1W_5+2W_2W_4- W_3W_3+5\p W_5-3\p (W_1W_4)\nm
&&\hskip 1cm +\p(W_2W_3)+4\p ^2 W_4-\p^2 (W_1W_3)+\p ^3 W_3 \nonumber .
\eea
Similar considerations can be done also for the $B_n$ algebras. 

\section{ Quantum considerations}

First we will show that the quantum generator has the same polynomial form 
as the classical one (\ref{clgen}) only the $c^{lm}_{ij}$ coefficients
get quantum corrections:
\be 
W^{C_n}_{2k}=\sum _{i,j,l,m} c^{lm}_{ij}(\k)\p^iW_l\p^jW_m  ,
\label{qgen}
\ee
where now $W_l$ denotes the explicitly known 
 quantised $A_n$-type expression, (\ref{anq},\ref{nwcn}) and normal 
ordering has to be understood.

The idea of the proof is to show that terms containing more then 
two $h_n$-s can be removed systematically since they are absent at 
the classical level, (\ref{wcn}). As a starting point
we write down the most general expression which commute with the first $(n-1)$ 
screening operators:
\be 
W^{C_n}_{2k}=\sum _{j_1, \dots , j_l;i_1,\dots ,i_l} 
c^{j_1,\dots, j_l}_{i_1, \dots , i_l}(\k)\p^{i_1}W_{j_1}\dots 
\p^{i_l}W_{j_l} .
\label{tr0}
\ee
Now we demand for it to be in the kernel of $Q_n$ ie. to contain 
$L^{(n)}=h_nh_n-2h_nh_{n-1}-(1-2\k)\p h_n$:
\be 
W^{C_n}_{2k}=\sum _{i_1,\dots i_l} 
b^{i_1,\dots i_l}(\k)\p^{i_1}L^{(n)}\dots \p^{i_l}L^{(n)} \{ {\rm terms \  
without \  }h_n 
\}  .
\label{tr1}
\ee
We have to redefine this $W$-generator such a way that the new one
contains the products of at most two $W_l$-s. We will do it inductively:
we define a partial ordering on the space of states. We say that
$\p ^{i_1}h_n \dots \p ^{i_k}h_n\p ^{i_{k+1}}h_{l_1}\dots
\p ^{i_{k+s}}h_{l_s}$ is bigger than $\p ^{j_1}h_n \dots \p ^{j_m}h_n\p
 ^{j_{m+1}}h_{p_1}\dots \p ^{j_{m+t}}h_{p_t} \!$ if $k>m$ or if $k=m$ then if
$s>t$. This induces an ordering among the $c$ coefficients if we associate 
the $h_nh_{n-1}\dots \p ^i h_{n-j+1}$ highest grade 
 term to $\p ^iW_j$. ( We note that the restriction of this map to the 
highest grade is injective, moreover normal ordering and other 
quantum corrections contribute at lower grades only). 
Now consider the highest grade terms in the expressions, (\ref{tr0},
\ref{tr1}), 
together with their $W_l$ and $L^{(n)}$ generators. We show in the
Appendix that there exist a classical $W$-generator with these highest
grade terms. This means however that  
the unneeded highest grade terms can be removed by redefining the 
$W$-generator, (\ref{tr0}). Doing this procedure 
from grade  to grade we end up with the original form (\ref{qgen}). 

In order to compute the coefficients explicitly we will use duality
\cite{KaWa2}. It states that if one replaces $h_i$ by $ \beta \tilde 
h_i$, where $\beta ^2= \k$ and defines  $\tilde W^{C_n}_{2k}(\beta)=
\beta ^{-2k}W^{C_n}_{2k}(\beta \tilde h_i)$, (and similarly for $B_n$)
 then 
\be
\tilde W_{2k}^{C_n}(\beta )=\tilde W_{2k}^{B_n}(-\beta ^{-1} ) .
\ee
Since the $A_n$ algebra is  self-dual this transformation acts only on 
the $c$ coefficients in the expression (\ref{qgen}). In more detail
this means that rescaling
the generators into their original forms the coefficient
of the $\k ^{i+j-m}$ term in $c^{kl}_{ij}(\k )$ for the $C_n$ algebra 
becomes the coefficient of the $(-\k) ^{m} $ term in the analoguous expression 
for $B_n$. 

The duality transformation makes it possible  to give the quantum generator 
at level two:
\be 
W^{C_n}_2=2W_2-\left (W_1\right )^2+(1-2\k)\p W_1 .
\ee
This form of the generator is $n$-independent and thus it is the same for 
all the \wbcn algebras.

Lets consider the next case, at level four:
\be
W^{C_n}_4
=aW_4+b\p W_3+cW_1W_3+W_2W_2+d\p W_1 W_2+eW_1\p W_2+f \p ^2 W_2 ,
\label{cn4}
\ee
where the terms without quantum corrections are:  $a=2$ and $c=-2$. 
Duality restricts the others as 
\be
b=3-4\k  \sp d=-1 \sp e=-(1-2\k) \sp f=1+ f_1\k +2\k ^2  ,
\ee
where the $f_1$ term is still unknown. However this is not surprising
since it may depend on the choice of the normal ordering of the 
$W_2W_2$ term, (all the other normal orderings are trivial in (\ref{cn4})).
 We note that the $W_k$ generators are exactly the same
for the  $A_n$, $B_n$ and $C_n$ algebras since the first $(n-1)$ screening
charges coincide. This means that computing the normal
ordering one can use any of these possibilities, moreover the normal ordering 
corresponding to the $N > n$ algebras. These choices are different since the
normal ordering  uses  the inverse of the inner product matrix.
The difference of the different orderings has to be in the kernel 
of the first $(n-1)$ generators so it can be expressed in terms of the 
$W_l$-s, moreover it contains at most two $h_n$-s, so it has
the form (\ref{qgen}). Note that the induced terms and so the quantised 
generator $W^{C_n}_4$  may
contain the classically removed $W_1$-s. However taking the $W_k \to 0 
\s k=2,3,4 $ limit the generator $W^{C_1}_4$ obtained has to be proportional
to $\p ^2 W_2^{C_1}$, so we can take such a linear  
combination of the generators $W^{C_n}_4$ and $\p ^2W^{C_n}_2$
 which does not contain the $W_1$ operator.
We chose a normal ordering which is exceptional in the 
sense that it contains the fewest new terms:
\be
:W_2W_2:=(W_2W_2)\vert _{C_n}+{\k \over 2}\p ^2 W_2-\dots  ,
\ee
where the dots mean that we removed the induced $n$-dependent $W_1$ terms 
and the subscript $C_n$ means that
we used the inner product matrix of the $C_n$ algebra. 
Now the term $f$ becomes $(1-\k)(1-2\k)$ and thus $f_1=-3\k$. 

Next consider the generator at level six. 
The terms which contain at most one derivatives are determined by 
duality:
\bea
&&2W_6-2W_1W_5+ 2W_2W_4- W_3 W_3+\p(W_2)W_3 +(5-6\k)\p W_5\nm
 &&-(3-4\k)\p (W_1)W_4-  (3-2\k) W_1\p W_4 +(1-2\k)W_2\p W_3  .
\label{I}
\eea
The other terms we compute in the $C_3$ model. Doing explicitly 
the calculations we have:
\bea
&-(1-\k)(1-2\k)W_1\p ^2W_3-(2-\k)(1-2\k)\p W_1\p W_3\nm
&-2(1-\k)(\p ^2 W_1) W_3+(1-\k)^2(1-2\k)\p ^3 W_3   .
\label{II}
\eea
If one computes the same terms in the $C_n$ case the coefficients may change 
due to the normal ordering which depends on $n$. The best one can do 
is to define an $n$-independent normal ordering. However to do it in
the general case one has to know the various defining
 relations of the $W$-algebras 
which is still missing. Concretely in the $C_3$ case there are 
induced lower level terms:
\bea
&{\k \over 12}(8\k-7)\p ^3 W_3-{\k \over 2}\p W_2\p W_2
-{\k\over 12}(10\k-7) W_1\p ^3 W_2\nm 
 &+{\k \over 4}(3-4\k)\p W_1\p ^2 W_2+{\k \over 2} 
 (2\k+1)\p ^2 W_1\p W_2+{\k \over 2}(12\k-7) \p ^3 W_1 W_2   .
\label{III}
\eea
The generator at level six  is the sum of (\ref{I}), (\ref{II}) 
and (\ref{III}). 

\section{ Conclusion }

We analyzed the generators of the \wbcn algebra using the explicitly 
know generators of the \wan algebra. Starting at the classical level 
we found a special  $n$-independent form 
which does not change under quantisation. In the general case to
 make it more explicit and go beyond the terms determined 
by duality one has to compute
the normal ordering explicitly for all the $W$-generators.

\bigskip
\bigskip

{\large {\bf Aknowledgements }}

\bigskip

We would like to thank L. Palla and L. Feh\'er for
the critical reading and the helpful comments.
This work was supported by OTKA T016251 and OTKA F019477. 

\section {Appendix}

The proof is very technical so we just sketch it. First we introduce
the notations:
In terms of the classical fields, $u_i=h_i-h_{i-1}$ (with $h_0=0$),
the sum of the leading terms of the generator $W_k$, which is denoted by $w_k$,
become the $k$-th elementary symmetric polynomial:
\be
w_k=\sum_{i_1<\dots <i_k}u_{i_1}\dots u_{i_k}
\ee
If $\Ai _{x}$ denotes the algebra of the differential polynomials
of the arbitrary classical field $x$ and we abbreviate $\Ai_{w_1,\dots , w_n}$
by $\Ai _w$ then the statement we have to prove is the following,
(see (\ref{tr0},\ref{tr1})),
\be
\Ai _{w}\cap \Ai _{u_n^2}=\Ai _{w_2^c,\dots, w^c_{2n}}=\Ai _{w^c}
\ee
where $w^c_{2k}$ denotes the sum of the leading terms of the classical generator
$W^C_{2k}$, ie. $w^c_2=w_1^2-2w_2, \dots, w^c_{2n}=(-1)^nw_nw_n$.

Clearly the algebra $\Ai _w$
is a subalgebra of $\Ai _{sym}$, the algebra of the symmetric differential
polynomials in the fields $u_i$-s for which a
polynomial basis is given by
\be
 p_{n_1,\dots, n_k}=\sum_{i_1,\dots, i_k; {\rm all} \neq}
\p^{n_1} u_{i_1}\dots  \p^{n_k} u_{i_k} ,\  n_i\leq n_{i+1}
 ; \ k=1,2,\dots , n
\ee
see \cite{Weyl} for the details. Consequently
\be
\Ai _{w}\cap \Ai _{u_n^2}=\Ai _{w}\cap \Ai _{sym^2},
\ee
where $\Ai _{sym^2}$ denotes the algebra of the symmetric differential
polynomials in the fields $u_i^2$-s, with polynomial basis
$P_{n_1,\dots ,n_k}=p_{n_1,\dots ,n_k}(u_i\to u_i^2)$. Note that
$P_{n_1,\dots ,n_k}=w^c_{2k}$ if $n_i=0$ for all $i$.

Define the endomorphism $\vf $ on $\Ai_{sym}$ by
\be
\vf ( p_{n_1,\dots, n_k})=\left \{ \ba {ll} p_{n_1, \dots, n_k} &
\mbox{ if  $ n_1=n_2=\dots =n_{k-1}=0$} \\ 0 &\mbox { otherwise} \ea   \right .
\ee
It is not hard to see that $\vf :\Ai_w\to \Ai_{sym}/{\rm Ker} \ \vf $ is an
isomorphism.

Note also that $P_{n_1,\dots ,n_k}=\displaystyle{
\sum_{i_1,\dots, i_k; {\rm all} \neq}}
\p^{n_1} u_{i_1}^2\dots  \p^{n_k} u_{i_k}^2$ is a linear combination
of terms of the form $p_{i_1,\dots ,i_k}p_{i_{k+1},\dots ,i_{2k}}$,
where $i_1,\dots, i_{2k}$ is any of the permutations of the non-negative
integers ${j_1, \dots, j_k,n_1-j_1, \dots, n_k-j_k}$.
This shows that $P_{n_1,\dots ,n_k}\in {\rm Ker }\  \vf$ except if
$n_1=n_2=\dots =n_{k-1}=0$. Since $ P_{0,0,\dots ,0,n_k}=\p^{n_k}
P_{0,0, \dots, 0}$ on $\Ai_{sym}/{\rm Ker}\vf $, which is isomorphic to
$\Ai_{w}$,  and $P_{0,0, \dots, 0}=
w^c_{2k}$ the statement is proved.

\end{document}